\documentclass[aps,prl,reprint,twocolumn,longbibliography]{revtex4-1}
\usepackage{graphicx}
\usepackage{amsmath}
\usepackage{amssymb}
\usepackage{comment}
\usepackage[colorlinks, allcolors=blue]{hyperref}
\usepackage[all]{hypcap}
\usepackage[mathlines]{lineno}

\newcommand{\pref}[2]{\hyperref[#1]{\ref{#1}(#2)}}
\newcommand{\preff}[2]{\hyperref[#1]{\ref{#1}#2}}
\newcommand{\eqpref}[1]{\hyperref[#1]{(\ref{#1})}}

\newcommand{\squig}{{\raise.17ex\hbox{$\scriptstyle\sim$}}}

\newcommand{\ket}[1]{| #1 \rangle}

\begin{document}
\title{Engineering a flux-dependent mobility edge in disordered zigzag chains}
\author{Fangzhao Alex An}
\author{Eric J. Meier}
\author{Bryce Gadway}
\email{bgadway@illinois.edu}
\affiliation{Department of Physics, University of Illinois at Urbana-Champaign, Urbana, IL 61801-3080, USA}
\date{\today}

\begin{abstract}
There has been great interest in realizing quantum simulators of charged particles in artificial gauge fields.
Here, we perform the first quantum simulation explorations of the combination of artificial gauge fields and disorder.
Using synthetic lattice techniques based on parametrically-coupled atomic momentum states, we engineer zigzag chains with a tunable homogeneous flux. The breaking of time-reversal symmetry by the applied flux leads to analogs of spin-orbit coupling and spin-momentum locking, which we observe directly through the chiral dynamics of atoms initialized to single lattice sites. We additionally introduce precisely controlled disorder in the site energy landscape, allowing us to explore the interplay of disorder and large effective magnetic fields.
The combination of correlated disorder and controlled intra- and inter-row tunneling in this system naturally supports energy-dependent localization, relating to a single-particle mobility edge. We measure the localization properties of the extremal eigenstates of this system, the ground state and the most-excited state, and demonstrate clear evidence for a flux-dependent mobility edge. These measurements constitute the first direct evidence for energy-dependent localization in a lower-dimensional system, as well as the first explorations of the combined influence of artificial gauge fields and engineered disorder. Moreover, we provide direct evidence for interaction shifts of the localization transitions for both low- and high-energy eigenstates in correlated disorder, relating to the presence of a many-body mobility edge. The unique combination of strong interactions, controlled disorder, and tunable artificial gauge fields present in this synthetic lattice system should enable myriad explorations into intriguing correlated transport phenomena.
\end{abstract}
\maketitle

\section{Introduction}

The idea that the transport of quantum particles in a random environment can be completely arrested due to the interference of multiple transport pathways was first pointed out by Anderson six decades ago~\cite{AndersonRandom}. While Anderson considered the localization of electrons in disordered solids, the presence of electron-phonon coupling and electron-electron interactions prohibit direct observation of most single-particle localization phenomena in such systems, even at low carrier density. In contrast, quantum simulation experiments using light~\cite{Segev2013} or atoms~\cite{SP2010} have become an important testbed for disorder physics, since in these systems the issues of lattice phonons and interparticle interactions are either naturally unimportant or can be precisely controlled. For cold atoms, the abilities to tune system dimensionality, applied disorder, atomic interactions, artificial gauge fields, and lattice geometry open up myriad possibilities for exploring novel localization phenomena.

In the absence of interactions, Anderson localization is the generic fate of quantum states in lower-dimensional ($d \leq 2$) systems featuring static, random potential energy landscapes and short-ranged tunneling~\cite{AndersonRandom,GangOfFour}. In higher dimensions, the increasing density of states with increasing energy ensures the possibility of delocalization. The exploration of an energy-dependent localization transition, i.e., a \emph{mobility edge}, has even been undertaken in atomic gases~\cite{Kondov-3D,Semeghini-MobEdge} in three dimensions through the precision control over disorder and atomic state energies. Cold atom techniques in principle also allow for the exploration of such physics in lower-dimensional systems, where mobility edges can be introduced by correlations in the applied disorder or modified lattice connectivities (e.g., through long-range tunneling).

Despite the exquisite control over cold atom systems and the observations of localization in one dimension (1D) over a decade ago, for both nearly random disorder~\cite{BillyAnd} and correlated pseudodisorder~\cite{Roati-Anderson}, single-particle mobility edges (SPMEs) in lower dimensions have gone unexplored. The reasons for this are somewhat technical -- it is quite difficult to modify lattice connectivities, and the varieties of engineered disorder that have been explored in experiment have either been practically random (speckle disorder~\cite{Lye-Speckle,BillyAnd,Kondov-3D,Semeghini-MobEdge}, with short-range correlations due to diffraction) or of a particular form of correlated disorder which, due to a peculiar fine-tuning, does not admit a SPME. In the latter case, the pseudo-disorder that arises in a lattice system due to shifts of the site energies by an added, weaker incommensurate lattice is well-described by the Aubry--Andr\'{e} model~\cite{AubAnd,Fallani-Glassy,Roati-Anderson,Schreiber-MBL}. While this form of correlated pseudodisorder allows for a localization transition in 1D, the fine-tuning of the cosine-distributed site energies and the cosine nearest-neighbor band dispersion results in an energy-independent metal-insulator transition, and thus the absence of a SPME. By deviating from this fine-tuned condition, either by modifying the band dispersion~\cite{Biddle-Long} or by modifying the form of the pseudodisorder~\cite{Ganeshan-GAA}, one can, in principle, controllably introduce a SPME in such a system.

\begin{figure}[t!]
	\includegraphics[width=\columnwidth]{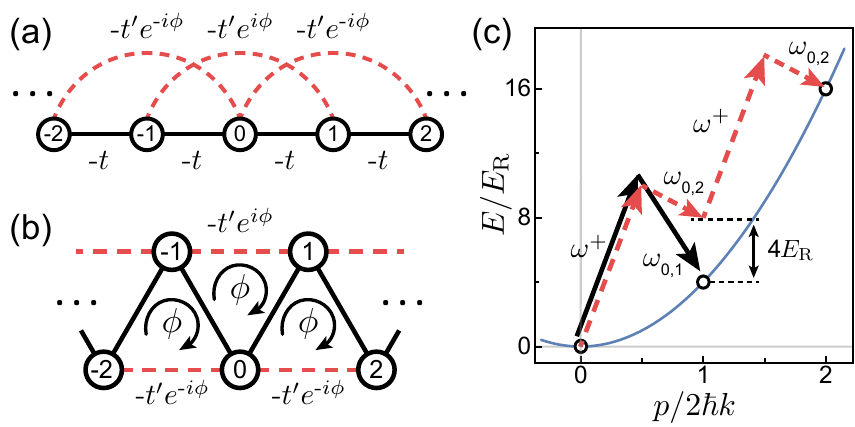}
	\caption{\label{FIG:fig1}
		\textbf{Constructing the zigzag lattice.}
		(a)~The one-dimensional multi-range hopping model with NN (solid black lines, $t$) and NNN (dashed red, $t'$) tunnelings.
		(b)~The two-dimensional zigzag lattice representation, formed by a rearrangement of the one-dimensional picture of (a). A uniform clockwise flux $\phi$ through each triangular plaquette is generated via NNN tunneling phases $\phi$ with alternating sign.
		(c)~Atomic dispersion indicating first- (black arrows) and second-order (dashed red arrows) Bragg transitions used to couple NN and NNN lattice sites, respectively, in the momentum-space lattice. The recoil energy is given by $E_R=\hbar^2k^2/2M_\text{Rb}$.
	}
\end{figure}

In this work, we add multi-ranged tunneling pathways to a one-dimensional lattice that features site energy pseudodisorder described by the Aubry--Andr\'{e} (AA) model. Specifically, we use our synthetic lattice system based on parametrically coupled atomic momentum states to engineer independently controllable nearest-neighbor (NN) and next-nearest-neighbor (NNN) tunneling terms (Fig.~\pref{FIG:fig1}{a}).
The combination of NN and NNN tunneling pathways results in closed tunneling loops that can support a nontrivial flux (Fig.~\pref{FIG:fig1}{b}), which we control directly through the complex phase of the various tunneling terms.
This system realizes an effective zigzag chain with a tunable magnetic flux. With the combination of controlled pseudodisorder and tunable flux, we perform the first explorations of the interplay of disorder and artificial gauge fields.

We observe direct evidence for a flux-dependent SPME in this system, through measurement of the localization properties of the extremal energy eigenstates. In addition to the SPME that results from multi-ranged hopping, we observe asymmetric (with applied flux) localization behavior of the systems lowest-energy and highest-energy eigenstates, caused by the presence of effectively attractive interparticle interactions in the lattice of momentum states~\cite{Gadway-Inter}. The influence of interactions is even more strongly evident in the case of the 1D AA with only NN tunneling, where a drastic shift in the localization transition is observed between low- and high-energy eigenstates, corresponding to a mobility edge driven purely by inter-particle interactions.

\section{Experimental methods}
To experimentally engineer effective zigzag chains, which are equivalent to a lattice model with NN and NNN tunneling terms, we coherently couple an array of discrete atomic momentum states with both first- and second-order Bragg transitions, as depicted in Fig.~\pref{FIG:fig1}{c}. Starting with atoms from a stationary Bose--Einstein condensate (BEC) of $\squig 10^5$ $^{87}$Rb atoms, we apply a set of counter-propagating lattice laser beams with wavelength $\lambda = 1064$~nm, wavenumber $k = 2\pi/\lambda$, and frequency $\omega^+ = c/2\pi \lambda$, allowing for quantized momentum transfer to the atoms in units of $\pm 2\hbar k$. The parametric coupling of states separated in momentum by $2\hbar k$, which mimics NN tunneling, is realized by using a pair of acousto-optic modulators to write a controlled spectrum of frequency components onto one of the lattice beams. Starting with atoms at rest, the counter-propagating beams are able to couple the momentum states $p_n=2n\hbar k$ as synthetic lattice sites. For example, to create a NN tunneling link between adjacent momentum states $p=0$ and $p=2\hbar k$, a first-order Bragg resonance (solid black arrows in Fig.~\pref{FIG:fig1}{c}) is fulfilled by matching the photon energy difference of the two laser fields to the added kinetic energy of an atom moving with momentum $p=2\hbar k$. More generally, there exists a unique energy difference between any pair of adjacent states with momenta $p_n$ and $p_{n+1}$, owing to the quadratic free-particle dispersion. In this way, the multiple frequency tones imprinted onto the one Bragg laser field enable the simultaneous addressing of many Bragg resonances.

In this study, we introduce the novel capability to engineer multi-range tunneling through the simultaneous addressing of first- and second-order Bragg transitions, shown in Fig.~\pref{FIG:fig1}{c} as solid black and dashed red arrows, respectively. Because each of the spectral tones associated with a given NN or NNN tunneling term is unique, we are able to individually control each of the tunneling links in our synthetic lattice. Specifically, all of the site energies, tunneling amplitudes, and tunneling phases in our synthetic zigzag chains are individually controlled by the strength, phase, and frequency of a corresponding frequency component of the multi-frequency beam. For all of the studies described herein, a total of 21 synthetic lattice sites (momentum states) are coupled through first- and second-order Bragg transitions.

\begin{figure*}[t!]
	\includegraphics[width=\textwidth]{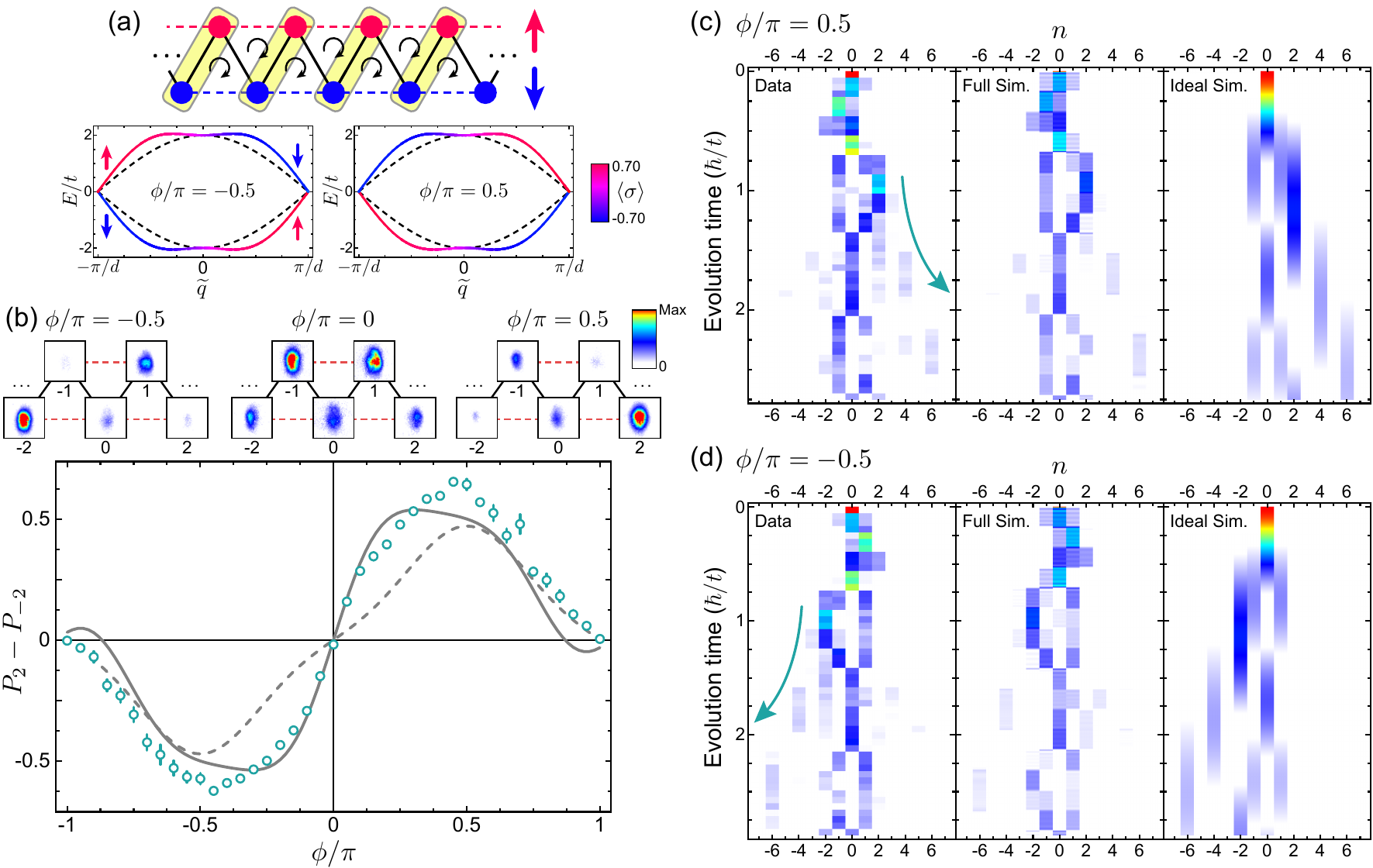}
	\caption{\label{FIG:fig2}
		\textbf{Chiral dynamics in the zigzag lattice.}
		(a)~Band structure for $\phi/\pi = \pm 0.5$ considering a two-site unit cell (yellow boxes in lattice cartoon), for tunneling ratio $t'/t = 0.62$. Color represents spin polarization $\langle \sigma \rangle$, or the overlap of the quasimomentum eigenstate with the top (red, spin up) or bottom (blue, spin down) row of the lattice. Dashed black curves represent the folded band structure for $t'/t= 0$. $\widetilde{q}$ should be considered ``quasiposition'' in our momentum-space lattice and is given in terms of the unit cell lattice spacing $d=4\hbar k$.
		(b)~Population imbalance between sites $2$ and $-2$ of the 21-site lattice, measured after 180 $\mu$s of dynamics ($\sim 1.05$ $\hbar/t$) with optical density (OD) images of atomic populations at $\phi/\pi = 0, \pm 0.5$ above. Dashed and solid curves represent an ideal simulation of the experiment using Eq.~\eqref{EQ:CleanHam} and a full simulation of experimental parameters, respectively.
		(c,d)~Site population dynamics for applied flux (c) $\phi/\pi= 0.5$ and (d) $\phi/\pi=-0.5$. Left to right: data, full simulation, and ideal simulation of experiment. Arrows indicate direction of chiral motion.
		Data for (b-d) were taken with averaged NN tunneling time $\hbar/t = 176(2)$ $\mu$s and tunneling ratio $t'/t = 0.622(3)$. All error bars denote one standard error of the mean. OD images in (b) and extracted site populations in (c,d) are plotted with the color scale in (b).
	}
\end{figure*}

In addition to local parameter control, this system supports site-resolved detection by a simple time-of-flight expansion period where the momentum states separate in space according to their momenta, after which absorption imaging is used to determine the population at each site.
A more detailed description of this momentum-space lattice scheme can be found in Refs.~\cite{Gadway-KSPACE,Meier-AtomOptics,Meier-TAI,ZZ-Supp}.

\section{Homogeneous gauge field studies}

We first demonstrate our control of a homogeneous synthetic gauge field in the zigzag lattice. We directly impose a synthetic magnetic flux $\phi$ on every three-site plaquette using engineered tunneling phases.
Because the plaquettes alternate pointing up and down, to generate a homogeneous positive flux $\phi$ we impose an alternating sign on the NNN tunneling phases, as shown in Fig.~\pref{FIG:fig1}{a,b}. The effective tight-binding Hamiltonian describing the 21-site zigzag lattice is then given by
\begin{equation}\begin{split}
\hat{H} = &-t\sum_{n=-10}^{9}\left(\hat{c}^{\dagger}_{n+1}\hat{c}_n+\text{h.c.}\right) \\
&-t'\sum_{n=-10}^{8}\left(e^{i(-1)^{n+1}\phi}\hat{c}^{\dagger}_{n+2} \hat{c}_n+\text{h.c.}\right),
\label{EQ:CleanHam}
\end{split}\end{equation}
where $t$ ($t'$) is the NN (NNN) tunneling energy and $\hat{c}^\dagger_n$ ($\hat{c}_n$) is the creation (annihilation) operator at site $n$.

The synthetic gauge field, which can lead to the breaking of time-reversal symmetry, allows us to engineer an analog of spin-momentum locking in the zigzag lattice~\cite{Atala-Chiral,Tai-IntFluxLadder,Mancini2015,Stuhl2015,Livi2016,An-FluxLadder,Kolkowitz2017}.
We consider the upper and lower rows of the lattice as an effective spin degree of freedom with (pseudo)spins $\sigma = 1$ and $-1$, respectively (Fig.~\pref{FIG:fig2}{a}). Under conditions of broken time-reversal symmetry ($\phi \neq 0,\pm\pi$) we expect to observe chiral trajectories for atoms ``polarized'' on one row of the lattice.
The band structure (shown for the tunneling ratio $t'/t=0.62$ used in experiment) of the lattice shows this correlation between the sign of the group velocity and the (colored) spin/row degree of freedom~\cite{ZZ-Supp}.
The two bands here reflect the two-site unit cell of the lattice, highlighted in yellow boxes.

To explore this spin-momentum locking in experiment, we initialize atoms on the lower row at site $0$ and quench on the tunnel couplings according to Eq.~\eqref{EQ:CleanHam}.
With zero applied flux, the population delocalizes across the lattice symmetrically, as shown in the top middle optical density (OD) image of Fig.~\pref{FIG:fig2}{b}.
For positive flux $\phi/\pi = +0.5$ (right panel), population initially in site 0 moves towards lattice site 2, corresponding to counter-clockwise chiral motion.
Under a negative flux $\phi/\pi=-0.5$ (left panel), population moves in a clockwise fashion to lattice site $-2$. These observed chiral flows for $\phi = \pm \pi/2$ are clear signatures of spin-momentum locking.

By tuning the applied flux, we map out the entire range of chiral behavior, as shown in Fig.~\preff{FIG:fig2}{(b), bottom}. Here we plot the population imbalance $P_2 - P_{-2}$ between lattice sites $2$ and $-2$, such that a positive (negative) value of imbalance indicates counter-clockwise (clockwise) motion. The data agree qualitatively with an ideal simulation of the experiment using only Eq.~\eqref{EQ:CleanHam} (dashed curve), but agree more closely with a full simulation of the system parameters (solid curve) which considers the exact form of atomic coupling to the many laser frequency components, accounting for off-resonant Bragg couplings~\cite{ZZ-Supp}.

\begin{figure*}[t!]
	\includegraphics[width=455pt]{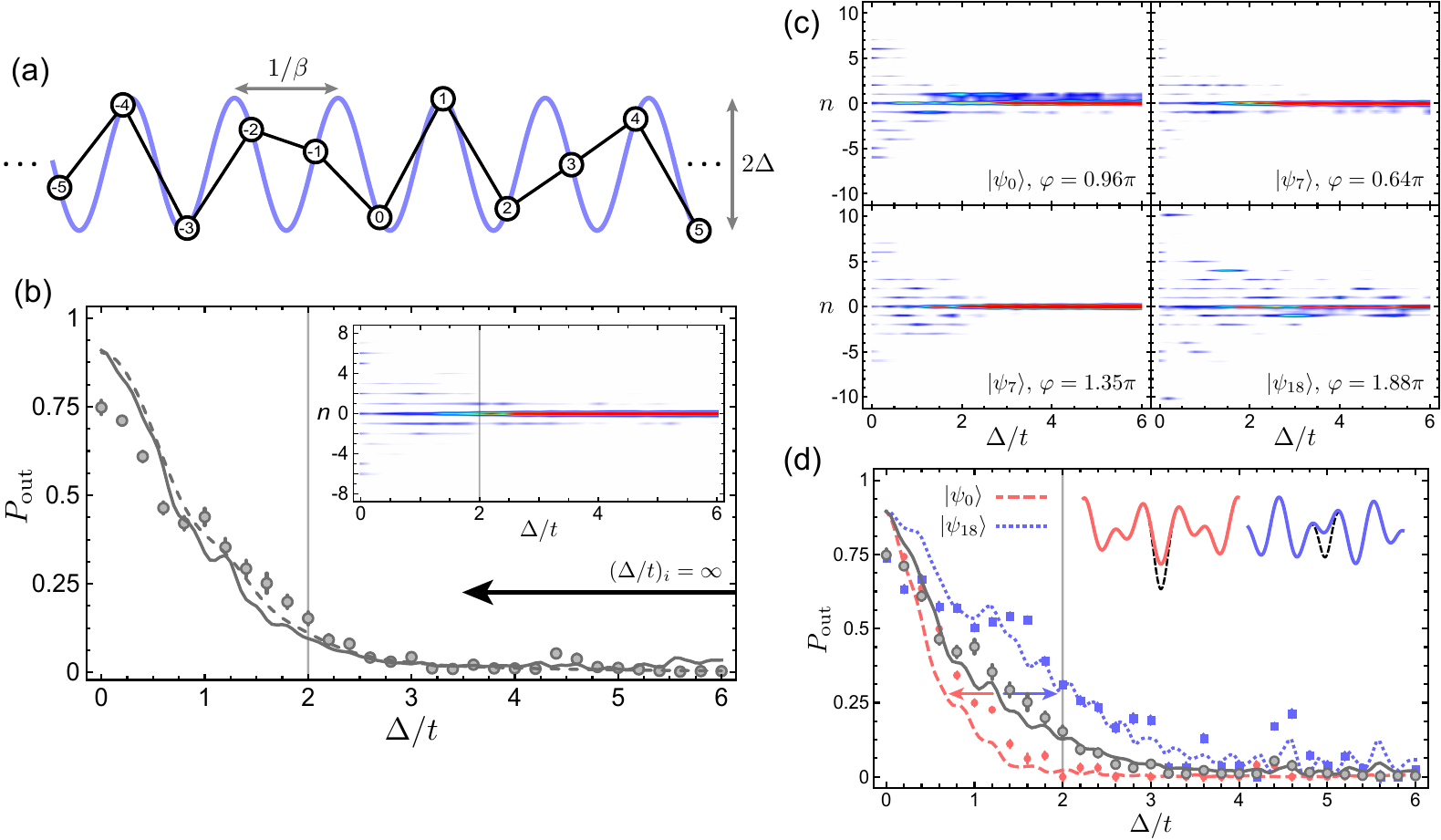}
	\caption{\label{FIG:fig3}
		\textbf{Localization in the 1D AA model with NN tunneling.}
		(a)~Lattice site energies under the AA model of quasiperiodic disorder, $\varepsilon_n = \Delta\cos\left(2\pi\beta n + \varphi\right)$, for periodicity $\beta = (\sqrt{5}-1)/2$ and amplitude $\Delta$.
		(b)~Measured population outside the central three lattice sites ($P_\text{out}$) vs disorder-to-tunneling ratio $\Delta/t$, averaged over four values of the AA phase $\varphi/\pi = \{0.96,0.64,1.35,1.88\}$. These $\varphi$ values correspond to the energy eigenstates $\{\ket{\psi_0},\ket{\psi_7},\ket{\psi_7},\ket{\psi_{18}}\}$, where $\ket{\psi_0}$ is the ground state and $\ket{\psi_{20}}$ is the highest excited state.
		Dashed and solid curves represent ideal and full simulations of experimental parameters, respectively.
		Arrow indicated experimental ramp from $(\Delta/t)_i=\infty$ to some final disorder value.
		Inset: Integrated OD image showing site populations vs disorder strength, also averaged over the four eigenstates.
		(c)~Integrated OD images for the individual eigenstates, labeled with the relevant AA phase value.
		(d)~$P_\text{out}$ vs disorder strength for eigenstates $\ket{\psi_0}$ (red circles) and $\ket{\psi_{18}}$ (blue squares).
		Gray circles are the averaged data of (c).
		Dashed red, dotted blue, and solid gray curves represent full simulation results including interactions of strength $U/\hbar \approx 2\pi \times 500$~Hz, for $\ket{\psi_0}$, $\ket{\psi_{18}}$, and the averaged data, respectively.
		Inset: Cartoon depicting lattice site energies for initial sites with low (left, red) and high (right, blue) energy. Dashed lines show the effect of attractive interactions on the initially-populated central site.
		Vertical lines in (b,d) indicate the critical disorder $(\Delta/t)_\text{c} =2$ for an infinite system without interactions.
		Error bars in (b,d) denote one standard error of the mean.
		OD images in (b,c) are plotted with the color scale in Fig.~\pref{FIG:fig2}{b}.
 	}
\end{figure*}

We are also able to directly observe the fully site-resolved chiral dynamics of initially localized atomic wave packets, as shown in Fig. \pref{FIG:fig2}{c,d}.
For positive flux, we see that atomic population moves counter-clockwise from site $0$ to site $2$, and further on to sites $4$ and $6$, remaining confined to the bottom row.
Because the initial state (site $0$) does not project entirely onto states with positive group velocity, a portion of the population stays near the center plaquette and oscillates between site $0$ and sites $\pm 1$.
Off-resonant Bragg coupling causes deviations from the ideal simulation (right), but these major qualitative features remain present in both the data (left) and full simulation (middle).
For the case of negative applied flux, we observe the opposite chiral behavior, demonstrating that the nature of the spin-momentum locking can be controlled by the applied synthetic flux.

\section{Localization studies}

Localization phenomena in disordered quantum systems depend intimately on the properties of applied disorder and on the connectivity between regions of similar energy. For random potential disorder in three dimensions, a localization-delocalization transition is assured for states with energies beyond a critical value due to an increasing density of states. For a given disorder strength, a mobility edge, or energy-dependent localization transition, is found in such a system~\cite{Kondov-3D,Semeghini-MobEdge}. In lower dimensions, for truly random potential disorder, all energy states remain localized in the thermodynamic limit even for arbitrarily small strengths of disorder~\cite{GangOfFour}.

Considering instead the influence of correlated pseudodisorder, one finds that the localization physics is strongly modified, with delocalization and mobility edges permitted even in lower dimensions. One form of quasiperiodic pseudodisorder that has been of interest to quantum simulation studies with both light~\cite{Lahini} and atoms~\cite{Roati-Anderson} is that described by the diagonal AA model. Interest in this model has stemmed in part from its intriguing localization phenomenology and connections to the Hofstadter lattice model~\cite{AubAnd,Thouless-AA,Hofst-1976}. Experimental interest in this form of disorder has also been driven by the relative ease of its realization through the overlap of two incommensurate optical lattices~\cite{Roati-Anderson}.

The AA model of pseudodisorder has interesting properties in the context of SPMEs. The highly correlated disorder allows for the possibility of a metallic, delocalized states in lower dimensions. However, a subtlety arises due to a correspondence between the distribution of pseudodisorder -- characterized by quasiperiodic, cosine-distributed site energies -- and the cosine dispersion in a NN-coupled 1D lattice. This fine tuning results in a metal-insulator transition that occurs at the same critical disorder value (in units of the tunneling energy) for all energy eigenstates, and thus the absence of a mobility edge. By moving away from this fine-tuned scenario in any number of ways -- by introducing longer-range hopping~\cite{Biddle-Long}, by modifying the pseudodisorder correlations~\cite{Ganeshan-GAA}, or by adding nonlinear interactions~\cite{DeisslerDis,Fallani-Glassy,Pas-Diso,Gadway-Glassy,Schreiber-MBL} -- a SPME can be introduced into the AA model.

The addition of longer-range tunneling, as in our zigzag lattice, allows for the band dispersion to be modified from its simple cosinusoidal form. For a flux of $\phi = \pm \pi/2$, as shown in Fig.~\pref{FIG:fig2}{a}, increasing the tunneling ratio $t'/t$ from zero leads to a deformation of the low-energy band structure from quadratic, to quartic, to forming a double-well structure~\cite{Gemlk-Shake,Lin-SOCBEC,Shaken-Chin}, with a symmetric modification of the band energies at high energy. The high ground state degeneracy of the quartic band in this system and of flat bands in similar multi-range hopping models has attracted great interest~\cite{Anisimovas-Zigzag,Eckardt-AntiFerro,Altman-sawtooth}. Such systems promise interesting localization properties under disorder~\cite{Biddle-Long}, and the inherent high single-particle degeneracy allows for the study of emergent physics driven by interactions~\cite{BoseLiquid,Anisimovas-Zigzag,Eckardt-AntiFerro,Altman-sawtooth}.
For all other flux values ($\phi \neq \pm \pi/2$) the dispersion of the bands at low and high energies is asymmetric, and this system permits the localization properties of the extremal energy eigenstates to be tuned through modification of the effective mass at low and high energies.

Here, we study the localization properties under the AA model on a 1D lattice and on the multi-range hopping zigzag lattice, observing evidence for an interaction-induced mobility edge as well as the emergence of a flux-dependent SPME.

\subsection{1D Aubry-Andr\'{e} localization transition}

We first examine the localization properties of the one-dimensional AA model, or the $t'/t=0$ limit of the zigzag lattice.
Figure~\pref{FIG:fig3}{a} shows this model's pseudodisordered distribution of site energies $\varepsilon_n = \Delta\cos\left(2\pi\beta n + \varphi\right)$, for an irrational periodicity $\beta = (\sqrt{5}-1)/2$ and a given value of the phase degree of freedom $\varphi$.
Under this model, all energy eigenstates experience a transition from delocalized metallic states to localized insulating states at the \emph{same} critical disorder, $(\Delta/t)_\text{c} = 2$, for an infinite system size.
To probe the crossover in our finite 21-site system, we initialize various energy eigenstates and explore their localization properties as a function of $\Delta/t$.

The experiment begins with population at site 0 (the BEC at rest) with all tunnelings turned off.
In this initial limit of infinite disorder ($\Delta/t)_i = \infty$, all eigenstates are trivially localized to individual sites of the lattice, with a vanishing localization length.
We can initialize our atoms in a particular energy eigenstate of the system through choice of $\varphi$, as the eigenstates and eigenstate energies are solely determined by the site energies in this $t=0$ limit.
We then slowly ramp the magnitude of the tunneling energy to a final value, and probe the localization properties of the prepared eigenstate as a function of $\Delta/t$. The ramp of $t$ to its final strength $t/\hbar=2\pi\times 1013(9)$~Hz (corresponding to a tunneling time of $\hbar/t=157(1)$~$\mu$s, determined through two-site Rabi oscillations) is linear and performed over 1~ms, slow enough to largely remain within the prepared eigenstate.
In each experiment, the disorder strength is fixed to a given value $\Delta$, such that the tunneling ramp (always to the same $t$ value) can be seen as traversing in parameter space from $\Delta/t = \infty$ to the chosen final value (shown as an arrow in Fig.~\pref{FIG:fig3}{b}).
We expect that for final values with $\Delta/t > (\Delta/t)_\text{c}$, the population should largely remain localized to the initial site, whereas for $\Delta/t < (\Delta/t)_\text{c}$ we should see population begin to delocalize across the lattice.

In Fig.~\pref{FIG:fig3}{b}, we plot the measured population outside the central three sites $P_\text{out}$, averaged over four realizations of the AA phase $\varphi/\pi=\{0.96,0.64,1.35,1.88\}$ corresponding to energy eigenstates $\{\ket{\psi_0},\ket{\psi_7},\ket{\psi_7},\ket{\psi_{18}}\}$, where $\ket{\psi_{0}}$ is the ground state and $\ket{\psi_{20}}$ is the highest excited state.
As expected, the measured delocalized fraction is almost entirely absent for large disorder, and grows steeply for $\Delta/t < (\Delta/t)_\text{c}$.
We find excellent agreement between our $\varphi$-averaged measurements and numerical simulation results based on our experimental ramp (dashed curve, idealized simulations ignoring off-resonant Bragg couplings) in Fig.~\pref{FIG:fig3}{b}, suggesting the observation of a localization crossover that is broadened due to finite-size effects as well as the finite ramp duration.
This same behavior can also be seen in the integrated optical density data, shown in the inset, which directly shows the averaged site populations for each final disorder value $\Delta/t$.
For large disorder, population remains localized to the initial site, while the metallic regime shows population spreading out to sites $n=\pm7$.

The data for individual energy eigenstates is also shown, both as integrated optical density images in Fig.~\pref{FIG:fig3}{c} and the $P_\text{out}$ observable in Fig.~\pref{FIG:fig3}{d}.
While all four data runs show localization crossovers, their positions in terms of a critical disorder-to-tunneling ratio $(\Delta/t)_\text{c}$ differ according to the state energies.
Visually, the ground state $\ket{\psi_0}$ appears to localize for smaller disorders than the intermediate energy eigenstates, with the highly excited state $\ket{\psi_{18}}$ requiring the largest critical disorder strength for localization.
While some of the broadening of the transition observed in Fig.~\pref{FIG:fig3}{b} can be attributed to effects of finite size and finite ramp durations, to a large degree it is explained by this averaging over unique localization transitions of different energy eigenstates.

The difference in localization properties for different energy eigenstates runs counter to our expectations of an energy-independent transition for the NN-coupled AA model, but can be explained by the presence of nonlinear atomic interactions in our momentum-space lattice~\cite{Gadway-Inter,ZZ-Supp}. In particular, the interactions between indistinguishable bosons in momentum space are effectively attractive and site-local, in the sense that direct interactions are present for collisions between two atoms occupying any pair of momentum modes, while exchange interactions are present only when two identical bosons occupy distinguishable modes~\cite{Bogoliubov-Rev,TrippJules}. Thus, while the momentum-space interactions are physically long-ranged and repulsive, they give rise to an effective local attraction. For atoms initially prepared at the site with lowest energy, attractive interactions can be seen to bring atoms further away from tunneling resonance with other sites (Fig.~\preff{FIG:fig3}{(d), inset}). Thus, such a state should remain localized even when the disorder drops below the single-particle critical value. In contrast, for atoms prepared at the highest energy site, attractive interactions effectively lower the total site energy and bring the atoms closer to tunneling resonance with the unoccupied lower-energy sites of the lattice (Fig.~\preff{FIG:fig3}{(d), inset}). Then, by filling the high-energy sites with attractively-interacting bosons, the disorder potential can be effectively smoothed out at high energies by atomic interactions~\cite{DeisslerDis}.

This behavior for our effectively attractive momentum-space interactions is exactly the opposite of that found for real-space repulsive interactions, the influence of which has previously been studied on ground state localization properties of the AA model~\cite{DeisslerDis}. The simulation curves in Fig.~\pref{FIG:fig3}{d} take into account the effective attractive interactions present in our system at an approximate, mean-field level (also ignoring the inhomogeneous atomic density and neglecting off-site contributions of the effective attraction, which arise due to partial indistinguishability of atoms in different momentum states resulting from superfluid screening~\cite{ZZ-Supp}). The simulations assume a mean-field interaction based on our condensate's central mean-field energy $U_0/\hbar \approx 2\pi\times 860$~Hz (as measured through Bragg spectroscopy), which is of the order of the single-particle tunneling energy $t/\hbar = 2\pi\times 1013(9)$~Hz.
To account for the inhomogeneous density distribution, we take a weighted average over homogeneous mean-field energies ranging from 0 to the peak mean-field energy $U_0$ to get an average mean-field energy of $U/\hbar \approx 2\pi \times 500$~Hz. We then use this average value as a homogeneous mean field energy in our simulations.
These simplified simulation curves already reproduce well the observed shifts of the localization transitions for the low- ($\ket{\psi_0}$) and high-energy ($\ket{\psi_{18}}$) states. These direct observations of interaction-induced localization and delocalization for low and high-energy states, respectively, are indicative of a many-body mobility edge. Such measurements are enabled by our unique ability to stably prepare \emph{any} particular eigenstate in our synthetic lattice.

\begin{figure*}[t!]
	\includegraphics[width=\textwidth]{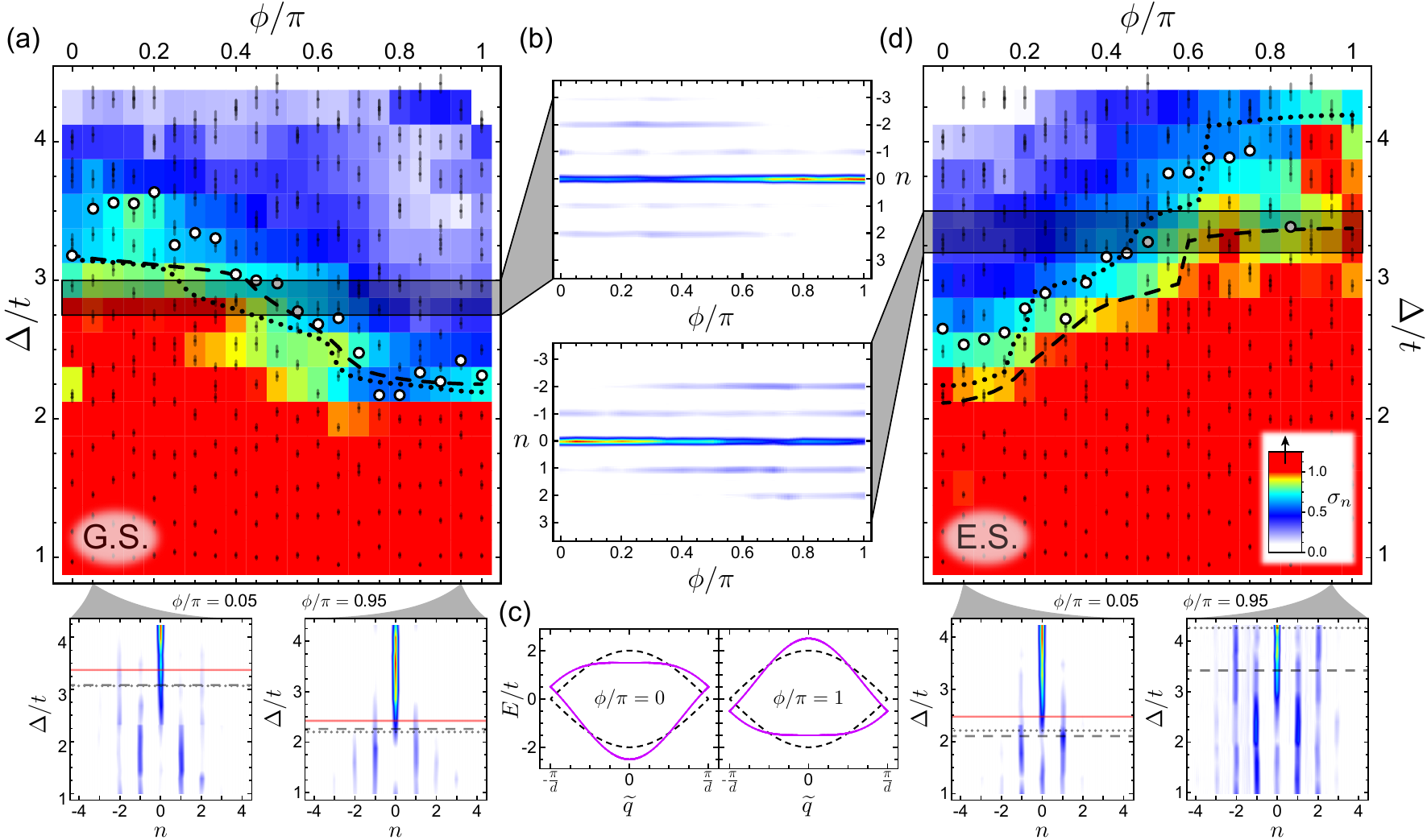}
	\caption{\label{FIG:fig4}
		\textbf{Flux-dependent localization transition in the AA model with multi-range tunneling.}
		(a)~Top: Ground state (GS) localization behavior for varying values of disorder $\Delta/t$ and flux $\phi/\pi$, in a zigzag lattice with NN tunneling $t/\hbar = 493(2)$~Hz and NN to NNN tunneling ratio $t'/t = 0.247(4)$. Colors indicate the width (standard deviation) of the site population distribution $\sigma_n$ (inset color scale in (d)), interpolated from the sampled data points (small solid black circles). The empirically-determined transition disorder strength between delocalized and localized regions is shown for the data (white circles), a non-interacting simulation of the experiment (dashed black line), and a simulation including attractive interactions of strength $U/\hbar\approx 2\pi\times 500$~Hz (dotted black line).
		Bottom: Localization properties as visualized by the integrated 1D density patterns for roughly 0 and $\pi$ flux. The 1D atomic distributions are interpolated from integrated optical density (OD) images for $\phi/\pi=0.05$ (left) and $0.95$ (right), shown as a function of $\Delta/t$. Horizontal lines show the empirically-determined localization transition point for data (solid red), non-interacting simulation (dashed gray), and simulation including interactions (dotted gray).
		(b)~Cuts as a function of $\phi$ showing site populations interpolated from integrated OD images, shown for the GS (top) and the highest excited state (ES, bottom). These integrated OD plots are averaged from data taken in the ranges $2.75 \leq \Delta/t \leq 3$ and $3.2 \leq \Delta/t \leq 3.5$, respectively, as indicated by the shaded regions.
		(c)~Band structure diagrams for the zigzag lattice with applied flux $\phi/\pi=0$ (left) and $\phi/\pi=1$ (right). As in Fig.~\pref{FIG:fig2}{a}, color represents spin polarization. Dashed black curves represent the folded band structure for $t'/t = 0$.
		(d)~ES localization behavior for varying disorder and flux values, with the same format as in (a). Bottom: Localization properties of the ES as a function of $\Delta/t$, also with the same format as in (a).
		Error bars in (a,d) denote one standard error of the mean.
		OD images in (a,b,d) are plotted with the color scale in Fig.~\pref{FIG:fig2}{b}.
 	}
\end{figure*}

\subsection{Localization studies in zigzag chains}

With the addition of longer-range tunneling, the energy-independent transition of the simple 1D AA model begins to depend critically on the eigenstate energy even at the single-particle level. By tuning the NNN tunneling strength and the artificial flux in our effective zigzag chains, we can introduce a tunable SPME through band structure engineering. While in the demonstration of control over flux and the observation of spin-momentum-locking in Fig.~\ref{FIG:fig2} we employed a tunneling ratio of $t'/t \approx 0.6$, here we work at a smaller value of $t'/t \approx 1/4$. Under this condition, a maximal difference in the band dispersion at low and high energies appears for flux values of 0 and $\pi$, where a quartic dispersion appears at high and low energies, respectively.

To probe the mobility edge, we prepare the two extremal energy eigenstates of the system, the ground state (GS) and the highest excited state (ES), and compare their localization properties.
As in the 1D study, our experiment begins with all atomic population prepared at site 0 with all tunnelings turned off, i.e., in the infinite-disorder limit of the system ($\Delta/t = \Delta/t' = \infty$) where all energy eigenstates are localized to individual sites of the lattice.
To initialize the atoms in a particular energy eigenstate of the system, we simply vary the AA phase: $\varphi = 0$ for the GS and $\varphi=\pi$ for the ES.

In short, we track how the prepared eigenstate evolves as the parameters of the Hamiltonian, given by
\begin{equation}\begin{split}
\hat{H}(V) \approx &\sum_{n=-10}^{10}\varepsilon_n'(V)\hat{c}^\dagger_n\hat{c}_n
-t\sum_{n=-10}^{9}\left(\hat{c}^{\dagger}_{n+1}\hat{c}_n+\text{h.c.}\right) \\
&-t'\sum_{n=-10}^{8}\left(e^{i(-1)^{n+1}\phi}\hat{c}^{\dagger}_{n+2} \hat{c}_n+\text{h.c.}\right) ,
\label{EQ:DisHam}
\end{split}\end{equation}
are smoothly and slowly varied to some final desired conditions of $\Delta/t$ for fixed tunneling ratio $t'/t$ and fixed flux $\phi$.
To help ensure adiabaticity over a large part of the parameter ramp, an extra potential offset of strength $V$ is added at the initial site $n=0$, such that the modified site energies are given by $\varepsilon_n'(V) = \Delta\cos\left(2\pi\beta n + \varphi\right)-V\delta_{n,0}$.
By setting $V > 0$ ($V < 0$) for the GS (ES), we further separate the initial eigenstate from the rest of the spectrum by a potential well (hill).
Starting from the initial limit of $V/t = \infty$ and $\Delta/t = \infty$, we adiabatically load our desired eigenstate by linearly ramping up both tunneling terms ($t$ and $t'$) over 2~ms while also smoothly removing the potential well by ramping $V$ to zero~\cite{ZZ-Supp}.

We perform this procedure over parameter ranges $1 \leq \Delta/t \leq 4.25$ and $0 \leq \phi/\pi \leq 1$, mapping out the localization behavior of the GS and the ES in Fig.~\pref{FIG:fig4}{a,d}.
We plot the standard deviation of the population distribution in the lattice, $\sigma_n$ (i.e., the momentum standard deviation $\sigma_p$ normalized to the spacing between sites of $2\hbar k$), where the values are resampled from the actual $(\Delta/t, \phi/\pi)$ points where data were taken (small black dots).
The $\Delta/t$ values of the data have variations and uncertainties stemming from variations and measured uncertainties in calibrated tunneling rates for the experimental runs, with an overall averaged NN tunneling rate $t/\hbar = 493(2)$~Hz and tunneling ratio $t'/t = 0.247(4)$.

For the ground state in Fig.~\pref{FIG:fig4}{a}, we see that the region of metallic, delocalized states (red region, corresponding to states with large $\sigma_n$) extends out to larger $\Delta/t$ values when the applied flux is near zero than for the case of an applied $\pi$ flux. This can also be seen in the integrated optical density images at bottom: sites as far as $n=\pm 2$ remain populated even at large disorder $\Delta/t \sim 3.5$ at small flux $\phi/\pi = 0.05$ (left), while for large flux $\phi/\pi = 0.95$ (right) population fully localizes for $\Delta/t > 3$. The top panel of Fig.~\pref{FIG:fig4}{b} highlights that for a fixed disorder-to-tunneling ratio of $\Delta/t \sim 2.9$, the GS can be driven from metallic to insulating by changing only the flux.

In the absence of flux, the shift of the GS localization transition to larger disorder values as compared to the $t'=0$ case is intuitive: simply adding longer-range tunneling increases the connectivity of the lattice, increasing the dispersion at low energy, and enhancing delocalization. As non-zero flux is added, however, the GS localization transition shifts towards smaller critical disorder values.
This effect is perhaps surprising when considering effects such as the suppression of weak localization by broken time-reversal symmetry, as observed recently in measurements of coherent back scattering~\cite{Muller-CBS}. However, in the context of our zigzag flux chains, this flux-enhanced localization of the GS is easy to interpret. The shift of the GS localization transition towards smaller $(\Delta/t)_\text{c}$ is driven by a flattening of the low-energy band dispersion, owing to kinetic frustration of the different tunneling pathways. The system is maximally frustrated at $\phi = \pi$ for $t'/t = 1/4$, corresponding to a nearly flat, quartic low-energy dispersion (Fig.~\preff{FIG:fig4}{(c), right}). Under these conditions, the states at low energy become heavy (large effective mass) and easier to localize in the presence of disorder.

In considering the flux-dependent localization properties of the highest energy eigenstate, a similar line of argumentation holds, but with the opposite trend with applied flux. The high energy states of the band structure are maximally dispersive for $\phi = \pi$, becoming flatter for decreasing flux, with a quartic band appearing for zero flux. The consequence of this modified band structure on the localization properties of the ES is reflected in the measured dependence of the ES localization properties following the parameter ramp to final $\Delta/t$ values for different flux values (Fig.~\pref{FIG:fig4}{d}). The flux-dependence of the localization transition is also seen in striking fashion in the integrated OD images at the bottom of Fig.~\pref{FIG:fig4}{d}: while the low flux panel clearly shows a transition from localized to delocalized behavior at $\Delta/t \sim 2.3$, in the high flux panel the site populations remain delocalized for all investigated disorder values.

\begin{figure}[t!]
	\includegraphics[width=\columnwidth]{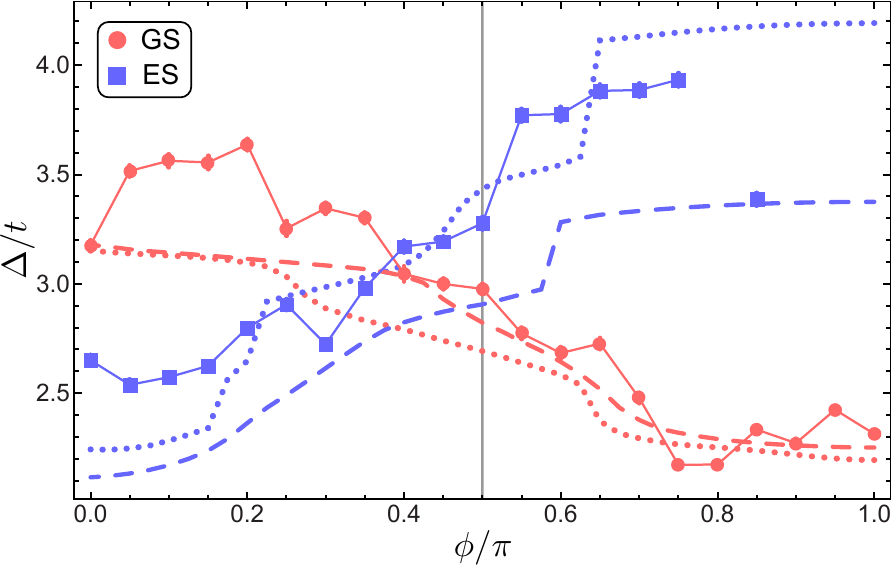}
	\caption{\label{FIG:fig5}
		\textbf{Flux-dependent mobility edge.}
		Empirically determined critical disorder-to-tunneling strength ratios marking localization transition for ground state (GS, red circles) and highest excited state (ES, blue squares), as shown separately in Fig.~\pref{FIG:fig4}{a,d}.
		Non-interacting and interacting simulations ($U/\hbar = 2\pi\times 500$~Hz used in the latter) are shown as dashed and dotted lines, respectively.
		For flux values where no critical disorder is plotted, atomic population was determined to be delocalized (based on the set threshold value of the standard deviation) over the full range of disorder strengths.
		Vertical gray line at $\phi/\pi=0.5$ denotes flux value at which the GS and ES curves should cross in the absence of interactions and any off-resonant coupling terms.
		Error bars denote one standard error of the mean.
 	}
\end{figure}

For both states, we empirically estimate the approximate ``critical'' disorder strength (normalized to $t$) relating to the metal-insulator transition by finding the $\Delta/t$ value at which $\sigma_n$ equals 0.68 lattice sites. This estimate is determined for each flux value of the data, and the extracted critical disorder strengths are shown as white circles in Fig.~\pref{FIG:fig4}{a,d}. We can compare these experimentally-extracted points to the predicted threshold values of disorder, based on numerical simulations of our experimental ramp protocol. These single-particle predictions are shown as dashed lines in Fig.~\pref{FIG:fig4}{a,d}, and show the same qualitative trend as the experimental points for both the GS and ES.

To better contrast the localization behavior of the GS and ES, we additionally plot both the experimentally determined transition points and the theory predictions for both extremal eigenstates together in Fig.~\ref{FIG:fig5}. With the two datasets overlaid, one can more clearly see the direct evidence for a flux-dependent SPME. While this sampling of the two extremal eigenstates does not determine the critical energy at which delocalization occurs for given values of $\Delta/t$ and $\phi$, it does provide the first direct experimental evidence for a SPME in lower dimensions.

The behavior of the transition $\Delta/t$ values for the GS and ES are nearly opposite to one another. For flux values near zero, the disorder strength needed to localize the GS exceeds that of the ES by nearly $t$, due to kinetic frustration of the high energy states. The situation reverses for flux values near $\pi$: the GS becomes localized at lower disorder strengths $\Delta/t \sim 2.3$, and the ES remains delocalized even up to the highest disorder value used in experiment $\Delta/t \sim 4.25$. This apparent asymmetry, i.e., that a larger magnitude of shift between the GS and ES transition points is found for flux values near $\pi$ than for flux values near $0$, is in disagreement with the single-particle prediction. Moreover, at the single-particle level the flux-dependence of the GS and ES localization properties should essentially be mirror images of one another (dashed lines, with a slight asymmetry resulting from effects due to off-resonant driving), such that their transitions points should cross very near to $\phi/\pi = 0.5$  (vertical gray line in Fig.~\ref{FIG:fig5}). However, the apparent crossing point is offset to lower flux values by nearly $0.1\pi$. As in the previously discussed case of the 1D AA model with only NN interactions (Fig.~\pref{FIG:fig3}{c,d}), the nonlinear interactions present in our atomic system are largely responsible for this asymmetry observed between the localization properties of our low and high energy eigenstates.

As described earlier in the context of the NN-coupled AA model, we can approximately capture the influence of the momentum-space interactions in this system by including a site-local mean-field attraction in a multi-site nonlinear Schrodinger equation~\cite{ZZ-Supp}, with an interaction energy that is determined independently by calibration via Bragg spectroscopy. Including these interactions (dotted lines, also shown in Fig.~\pref{FIG:fig4}{a,d}), the transition lines get shifted to lower (GS) and higher (ES) disorder values, so that they cross at lower flux values.
The interacting simulation results better capture the localization properties of the ES, which was shifted to significantly higher disorder strengths than was predicted at the single-particle level. It also qualitatively captures the shift of the crossing of the critical disorder curves in Fig.~\ref{FIG:fig5} to lower flux values, although it predicts a slightly larger shift than seen in experiment. In the future, by studying fluctuations of the atomic number distribution and inter-site correlations in our synthetic lattice, or by more closely studying fine features of the localization properties, this simulation platform may enable unique explorations into the physics of interacting disordered systems, in particular related to the physics of many-body localization. It also offers a unique platform to study the interplay of disorder, artificial gauge fields, and interactions.

\section{Conclusions}

This work represents the first direct observation of a single-particle mobility edge in lower dimensions, which is enabled by the unique ability to stably prepare atoms in any energy eigenstate and explore their localization properties in a system with precisely controlled disorder and tunable artificial gauge fields. We also present the first direct quantum simulation evidence for a many-body mobility edge, studied through a shift of the localization properties of low- and high-energy eigenstates in the 1D AA model that arise due to many-body interactions. These interaction shifts are also observed in the localization transitions of a multi-range hopping AA model that admits a flux-dependent SPME, leading to the interplay of single-particle and many-body shifts of the localization transition for states at different energies.

This work also constitutes the first quantum simulation study combining synthetic gauge fields and disorder, and its extension to fully two-dimensional lattices beyond coupled chains promises to pave the way towards studies of disordered quantum Hall systems. In particular, by moving to a larger system containing bulk lattice sites, a robustness of the observed chiral-propagating modes to disorder (similar to the robustness to disorder observed recently for the bulk winding of chiral symmetric wires~\cite{Meier-TAI}) should be readily observable.

\section{Acknowledgements}
We thank S. Hegde, K. Padavi\'{c}, and S. Vishveshwara for helpful discussions, and we thank M.~A. Highman and J. Ang'ong'a for helpful discussions and a careful reading of the manuscript.
This material is based upon work supported by the National Science Foundation under grant No.~1707731.

Following the posting of an earlier version of this manuscript, two related works were also posted, exploring the SPMEs~\cite{Bloch-SPME} that appear naturally due to NNN tunneling in weak real-space optical lattices featuring AA disorder~\cite{Boers-Bichromatic} and exploring the interplay of disorder and artificial gauge fields in kicked rotor systems~\cite{French-DisGauge}.


%

\end{document}


\title{Supplemental Materials for ``Engineering a flux-dependent mobility edge in disordered zigzag chains''}
\author{Fangzhao Alex An}
\author{Eric J. Meier}
\author{Bryce Gadway}
\email{bgadway@illinois.edu}
\affiliation{Department of Physics, University of Illinois at Urbana-Champaign, Urbana, IL 61801-3080, USA}
\date{\today}
\maketitle

\section{Experimental setup}

Our experiments begin with a Bose--Einstein condensate (BEC) containing \squig$10^5$~$^{87}$Rb atoms, held in an optical dipole trap primarily formed by a single focused laser beam (wavelength $\lambda=1064$ nm, wavenumber $k=2\pi/\lambda$) with weak additional trapping provided by one other beam.
To create the lattice, we allow this primary beam to pass through two acousto-optic modulators (AOMs) which, together, write onto the beam a spectrum of radiofrequency tones in the sub-MHz range.
This multi-frequency beam is sent back towards the atoms along the same path as the incoming single-frequency beam.
The interference of these two beams creates a time-dependent optical potential comprised of multiple superimposed optical lattices moving at different velocities.
Each of these velocities is determined by the frequency difference between one frequency component of the multi-frequency beam and the incoming beam, specifically tuned to address a particular Bragg transition between two discrete momentum states.
The simultaneous driving of many Bragg transitions mimics tunneling between sites (momentum states) in our synthetic lattice~\cite{Gadway-KSPACE,Meier-AtomOptics}.

\begin{figure}[t!]
	\includegraphics[width=\columnwidth]{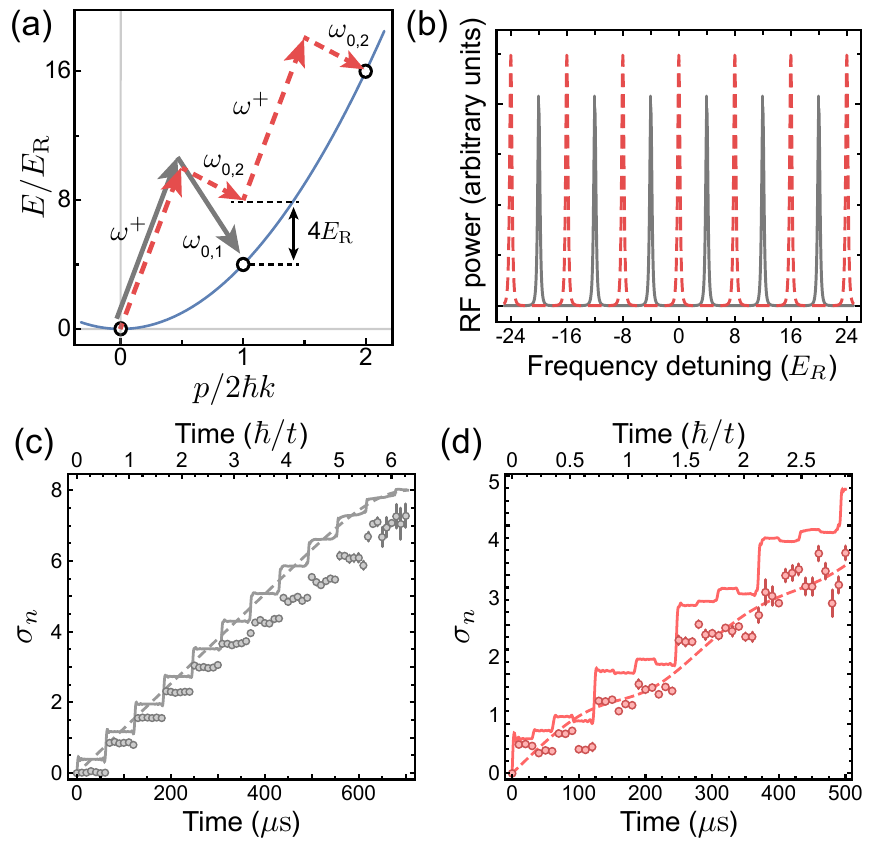}
	\caption{\textbf{Effects of off-resonant excitations.}
	(a)~Dispersion relation and depiction of Bragg transition pathways for NNs (solid grey) and NNNs (dashed red). Energy is shown in units of recoil energy, $E_R=\hbar^2k^2/2M_\text{Rb}$.
	(b)~Frequency spectrum (w.r.t. the incident beam frequency) used to create a 21-site lattice with 20 NN and 19 NNN links.
	(c)~Dynamics of the width of the population distribution, $\sigma_n$ (in units of synthetic lattice spacing), for a lattice with $t'/t=0$.
	(d)~Dynamics of $\sigma_n$ for a lattice with $t'/t = 0.628(6)$ and applied flux $\phi/\pi=-0.5$ (same data as in Fig.~2(d)). Dashed and solid curves in (c,d) represent results from an ideal simulation of the experiment and a full simulation accounting for off-resonant coupling, respectively.
	\label{FIG:figS1}}
\end{figure}

By varying the amplitude and phase of each frequency component, along with the frequency detuning from Bragg resonance, we can control the amplitude and phase of each effective tunneling link, as well as each site-energy term.
Because the single-particle dispersion relation is quadratic ($E = p^2/2M_\text{Rb}$), all of the first- and second-order Bragg transitions have unique frequencies and can be individually addressed.

Next-nearest-neighbor (NNN) tunnelings are realized via a four-photon, second-order Bragg process. Shown in Fig.~\pref{FIG:figS1}{a} as dashed red lines, this involves virtually absorbing two photons from the single frequency beam ($\omega^+$) and emitting two photons into the multi-frequency beam ($\omega_{0,2}$), or vice versa.
Control of the effective lattice parameters for these four-photon processes requires slightly different considerations compared to the nearest-neighbor (NN) terms.
For example, to enable a NNN tunneling phase of $\pi$, we apply half this phase ($\pi/2$, relative to incoming field) to the corresponding frequency component $\omega_{0,2}$. A similar consideration holds for the relationship between site energy and frequency detuning from resonance.
More generally, these differences can be summarized in the relationships between the tunneling terms for NN ($t_{nn} e^{i\varphi_{nn}}$) and NNN ($t_{nnn} e^{i\varphi_{nnn}}$) processes.
Taking into account the field strengths (assumed to be real) of the incoming beam ($\Omega_I$) and a particular frequency component of the multi-frequency beam ($\Omega_R$), the phases of these same fields ($\phi_I$ and $\phi_R$), the large single-photon detuning from atomic resonance $\Delta$ (relating to roughly 100~THz for our laser wavelength $\lambda=1064$~nm), and the recoil energy $E_R = h^2 /2 M_\text{Rb} \lambda^2$, these terms are given at resonance as:
\begin{equation}
\label{EQ:S1}
t_{nn} e^{i\varphi_{nn}} = \frac{\Omega_I \Omega_R}{\Delta} e^{i(\phi_R - \phi_I)} \ \ \  \mathrm{and}
\end{equation}
\begin{equation}
\label{EQ:S2}
t_{nnn} e^{i\varphi_{nnn}} = \frac{\Omega_I^2 \Omega_R^2}{\Delta^2 (4 E_R)} e^{i(2\phi_R - 2\phi_I)}\ .
\end{equation}

\section{Tunneling times and tunneling ratios}

Here we present the exact NN tunneling times $\hbar/t$ and NNN to NN tunneling ratios $t'/t$ for each individual data set shown in the main text.
For the variable flux data of Fig.~2(b), $\hbar/t=172(1)$~$\mu$s and $t'/t = 0.633(3)$.
For the dynamics under $\phi / \pi = 0.5$ of Fig.~2(c), $\hbar/t=182(1)$~$\mu$s and $t'/t = 0.605(5)$.
For the dynamics under $\phi / \pi = -0.5$ of Fig.~2(d), $\hbar/t=174(2)$~$\mu$s and $t'/t = 0.628(6)$.
The tunneling ratio used to make the band structures of Fig.~2(a) was an average of these three values: $t'/t = 0.622(3)$.
For the 1D Aubry-Andr\'{e} data of Fig.~3, $\hbar/t=157(1)$~$\mu$s.
For the multi-range hopping Aubry-Andr\'{e} data of Fig.~4 and Fig.~5, $\hbar/t=323(1)$~$\mu$s and $t'/t = 0.247(4)$, averaged over all data points taken.

\section{Off-resonant excitations}

While we seek to address individual Bragg transitions with single frequency components, in practice we apply the full spectrum of frequencies to the condensate, as shown in Fig.~\pref{FIG:figS1}{b}.
Thus each transition is not only addressed by one resonant frequency, but also feels the effects of all of the other non-resonant frequency components.
For a lattice with only NN tunnelings, the frequency components are equally spaced at $8E_R/\hbar = 8\times\hbar k^2/2M_\text{Rb} \approx 2\pi \times 16.2$~kHz.
Adding NNN links halves the spacing of applied frequency components to $4E_R/\hbar \approx 2\pi\times 8.1$~kHz.

On a lattice with only NN tunnelings, the off-resonant couplings result in step-like intervals in the dynamics (Fig.~\pref{FIG:figS1}{c}).
For this data (from Ref.~\cite{An-Disorder}), atomic population undergoes a continuous-time quantum walk on a 21-site lattice engineered with 20 equally-spaced frequency teeth.
Due to the equal spacing of the frequency teeth, the off-resonant effects add up constructively. This is evident in the period $T$ between these steps, which corresponds exactly to the spacing between adjacent frequency teeth, $T = h / 8E_R$. We have shown in a previous study~\cite{An-Disorder} that adding random tunneling phases onto equally-spaced frequency teeth suppresses these steps, resulting in smoother dynamics.

We note that the magnitude of the tunneling rate plays a significant role in the magnitude of off-resonant effects. If the tunneling rate is comparable to or greater than the frequency spacing between the first-order Bragg resonances ($t/\hbar \gtrsim 8E_R/\hbar$), then each frequency tooth may address multiple transitions. The extreme limit of non-resonant addressing is often encountered in cold atom experiments, e.g., when a deep stationary potential (two interfering fields with equal frequency) that is suddenly turned on results in Kapitza--Dirac diffraction~\cite{Gould-KD} of atomic matter waves. With respect to synthetic lattices, this can be viewed as a system with constant NN tunneling terms in the presence of a quadratic potential, set simply by the single-particle dispersion relation. This leads to expansion dynamics in momentum space at short times up until population reaches outer regions where the site (momentum state) energy roughly equals the effective tunneling bandwidth~\cite{Huckans-Collapse}.

On the other hand, in the limit where the tunneling energy is much smaller than the energy spacing between relevant Bragg resonances (i.e., in the limit that the rotating wave approximation is valid), each transition will be ideally addressed by only a single spectral component, and the step-like behavior in Fig.~\pref{FIG:figS1}{c} (data, solid curve) should approach the ideal smooth behavior (dashed curve). Intuitively, this occurs as the number of steps per tunneling period gets very large, or in other words as the tunneling time $\hbar/t$ gets much larger than the step period $T$, such that the dynamics are spread out over many, many steps.
For the data in Fig.~\pref{FIG:figS1}{c}, we are in the intermediate regime (with a tunneling time $\hbar/t = 111.6(7)$~$\mu$s, corresponding to a tunneling rate of $t/\hbar = 2\pi\times 1425(9)$~Hz), where off-resonant coupling primarily results in the observed dynamics with small step-like behavior.

For the zigzag lattice, we introduce NNN frequency teeth that halve the spacing to $4E_R/\hbar\approx 2\pi\times 8.1$ kHz (addition of dashed red peaks in Fig.~\pref{FIG:figS1}{b}). This results in longer steps of exactly twice the duration, i.e., $T' = 2T$ (Fig.~\pref{FIG:figS1}{d}) due to off-resonant first-order Bragg processes. The smaller structure on top of these long steps, with a spacing of $T/2$ relating to a frequency spacing of $16E_R/\hbar$, is due to off-resonant second-order Bragg processes. For this data, we used a tunneling ratio $t'/t = 0.628(6)$ and NN tunneling time $\hbar/t = 174(2)$ $\mu$s, corresponding to a tunneling rate $t/\hbar = 2\pi \times 917(9)$~Hz.

The ``full'' simulations in Fig.~\pref{FIG:figS1}{c,d} (solid curves) and in the main text account for both resonant and off-resonant driving on every Bragg transition, and thus retain these step-like features.
The ``ideal'' simulations in Fig.~\pref{FIG:figS1}{c,d} (dashed curves) and in the main text ignore off-resonant effects and consider only the smooth behavior of the idealized tight-binding Hamiltonians.

\section{Band structure calculations}

The band diagrams shown in Fig.~2(a) and Fig.~4(c) were calculated using the same method as described in Sec. II of Ref.~\cite{Anisimovas-Zigzag}.
While the studies there focused on a semisynthetic lattice with one synthetic dimension and one real-space dimension, the same physics hold for our fully synthetic momentum-space lattice.

We consider the zigzag lattice in the absence of any applied disorder. We can take the rows of the lattice to be an effective spin degree of freedom, with a two-site unit cell comprised of one spin up ($\sigma = +1$) site from the top row and one spin down ($\sigma = -1$) site from the bottom row.
We generate the spinful dispersion by calculating the $2\times 2$ Hamiltonian introduced in Ref.~\cite{Anisimovas-Zigzag} at each value of quasimomentum.\footnote{We note that because our lattice is in momentum space, the quasimomentum $\qs$ should really be called ``quasiposition,'' though we stay with the conventional term ``quasimomentum.''}
In terms of the creation and annihilation operators at spin $\sigma$ and quasimomentum $\qs$ ($\hat{c}^\dagger_{\sigma,\qs}$ and $\hat{c}_{\sigma,\qs}$), the Hamiltonian is given by
\begin{equation}
\label{EQ:S3}
\hat{H}=\sum_{\qs}
\left({\begin{array}{cc}
    \hat{c}^\dagger_{1,\qs} & \hat{c}^\dagger_{-1,\qs}
  \end{array}}
\right)
\left({\begin{array}{cc}
    h_{11} & h_{12} \\
    h_{21} & h_{22}
  \end{array}}
\right)
\left({\begin{array}{c}
    \hat{c}_{1,\qs} \\
		\hat{c}_{-1,\qs}
  \end{array}}
\right),
\end{equation}
where $h_{jj} = -2 t' \cos\left[\qs d + \phi(-1)^j\right]$ and $h_{12} = h_{21} = 2t\cos\left[\qs d/2\right]$ for lattice spacing $d=4\hbar k$ (not $2\hbar k$ due to the two-site unit cell). By diagonalizing this Hamiltonian for every value of $\qs \in \left[-\pi/d,\pi/d\right]$, we generate the double-band dispersions. We note that for $t'/t=0$ (dashed black curves in Fig.~2(a) and Fig.~4(c)), the dispersion relation is cosinusoidal, but folded back at the edges of the Brillouin zone due to the two-site unit cell.

The spin magnetization $\langle \sigma \rangle$ is simply the projection of the quasimomentum eigenvectors derived from Eq.~\eqref{EQ:S3} onto the rows of the lattice. We take the difference between the projections onto the upper row and the lower row such that a positive (negative) $\langle\sigma\rangle$ corresponds to population on the upper (lower) row.

\section{Influence of interactions}

As mentioned in the main text, atomic interactions show effects on the localization properties of both the 1D Aubry-Andr\'{e} data in Fig.~3 and the longer-range Aubry-Andr\'{e} data in Fig.~4 and Fig.~5.
Interactions in momentum-space lattices are described in detail in Ref.~\cite{Gadway-Inter}, where we show that effects like self-trapping can be observed when the mean-field energy becomes large compared to the tunneling.

Atoms in a particular momentum state experience an added positive self energy due to repulsive cold collisions, i.e., mean-field interactions.
In addition, atoms overlapped in space but occupying distinct spatial eigenstates (i.e., distinguishable plane-wave momentum states) experience both a direct interaction as well as an added exchange energy, resulting in twice as large a repulsive energy~\cite{Bogoliubov-Rev,TrippJules}.
For a fixed total density, this situation where atoms occupying the same momentum state have a weaker repulsive interaction energy may be recast as an effectively site-local attraction with a scale set by the mean-field interaction energy $U$. In reality, for an interacting degenerate Bose gas, superfluid screening can make distinct plane wave states partially indistinguishable, resulting in some off-site contribution to the effective attraction (although this vanishes as $U$ becomes much less than $2 E_R$).

As discussed in Ref.~\cite{Gadway-Inter}, these interactions shift the Bragg resonance frequencies away from the single particle resonances.
Under typical experimental conditions, we measured this shift to be $\squig 300$~Hz, relating to a peak mean-field energy $U_0 = gn_0 \approx 2\pi\hbar\times 860$~Hz at the center of the harmonic trap, and an homogeneous mean-field energy $U/\hbar \approx 2\pi \times 500$~Hz averaged over the entire trap (the measured shift is distinct from the average $U$ value due to a combination of the aforementioned screening effects and the long duration of the Bragg pulses used in this determination).
The central atomic density is $n_0 \approx 10^{14}~\text{cm}^{-3}$ and $g = 4\pi\hbar^2 a/M_\text{Rb}$, where $a$ is the scattering length~\cite{Stenger-Bragg}.

We incorporate this mean-field energy $U$ by considering an attractive interaction that depends on the population of atoms at each site~\cite{Gadway-Inter}, resulting in the curves shown in Fig.~3(d) and the dotted theory curves in Fig.~4. Specifically, the evolved state at the end of the tunneling-ramps described in the main text are found by solving a time-dependent multi-site nonlinear Schr\"{o}dinger equation that includes a local attractive self-nonlinearity $- U |\psi_n|^2$, where the $\psi_n$ are $c$-numbers (with normalization $\Sigma_n |\psi_n|^2 = 1$) relating to the atomic field terms at each site.
This approach is approximate in a number of ways: it ignores quantum fluctuations, ignores off-site contributions to the effective atomic interaction, ignores energy-dependent corrections to the collisional scattering cross-section, ignores spatial variations in the atomic density in our trapped sample, ignores effects such as the loss of spatial overlap of the momentum wavepackets, and explicitly restricts the collisions to be mode-preserving (ignoring both $s$-wave collisions that may scatter atoms out from the considered set of $21$ modes into many ``halos'' of additional states, as well as mode-changing collisions within our defined set of states, that would be energetically suppressed in the absence of our drive fields, but may be effectively enabled through higher-order, Bragg-mediated processes).
We point out that a fuller quantum treatment of the problem (still restricted to being mode-preserving, still ignoring spatial variations of the density, still ignoring loss of spatial overlap of momentum states, and still ignoring energy-dependent corrections to the scattering cross-section) would instead include an interaction term $(U/N)\Sigma_{n,n'}c^\dag_n c^\dag_{n'}c_n c_{n'}$ in the effective tight-binding Hamiltonians of the main text. Even for our modest 21-mode system, the time-dependence of this problem would become intractable for particle numbers well below the $\sim 10^5$ used in experiment.

\section{Ramp procedure}

For the localization studies of the zigzag lattice of Fig.~4 and 5, we slowly load into the extremal eigenstates (ground state or highest excited state) of the Hamiltonian we wish to explore (Eq.~(2)).
We begin, as described in the main text, by preparing with high fidelity the ground state of the system in the zero tunneling ($t=t'=0$) limit by shifting the Aubry-Andr\'{e} site energy distribution such that the initial site has the lowest energy ($\varphi = \pi$ for $\Delta > 0$).
To ensure that there is a relatively large energy gap from this initially populated ground state to all other eigenstates even after finite tunneling is introduced, we add an effective potential well of depth $V$ at the central site.\footnote{For the most excited state, we set the initial site to have the highest energy ($\varphi = 0$) and impose a potential hill of height $V$ at the initial site $n=0$.}
Then, over the course of 2~ms, we smoothly vary the system parameters until we reach the desired Hamiltonian.
If these ramps are quasistatic adiabatic, such that the energy associated with the ramp rate is much smaller than the smallest energy gap encountered, then this procedure should prepare the desired ground state with high fidelity.

First, we describe the different ramps used in experiment. The depth of the potential well at site $n=0$ is ramped from $V$ to zero over 2~ms (for comparison, the NN tunneling time for this experiment was $\hbar/t = 304(4)$~$\mu$s).
Over the same 2~ms duration, we also ramp both the NN and NNN tunneling amplitudes linearly from zero to their final magnitudes $t$ and $t'$, respectively.
Over the course of this ramp we preserve the flux distribution, imposed by fixed tunneling phases.
We additionally preserve the ratio of $t' / t$ by ramping the field strengths of the first- and second-order Bragg spectral components ($\propto \tau$ and $\propto \sqrt{\tau}$) according to their distinct scalings with the applied field strengths (Eqs.~\eqref{EQ:S1} and \eqref{EQ:S2}).

\begin{figure}[t!]
	\includegraphics[width=\columnwidth]{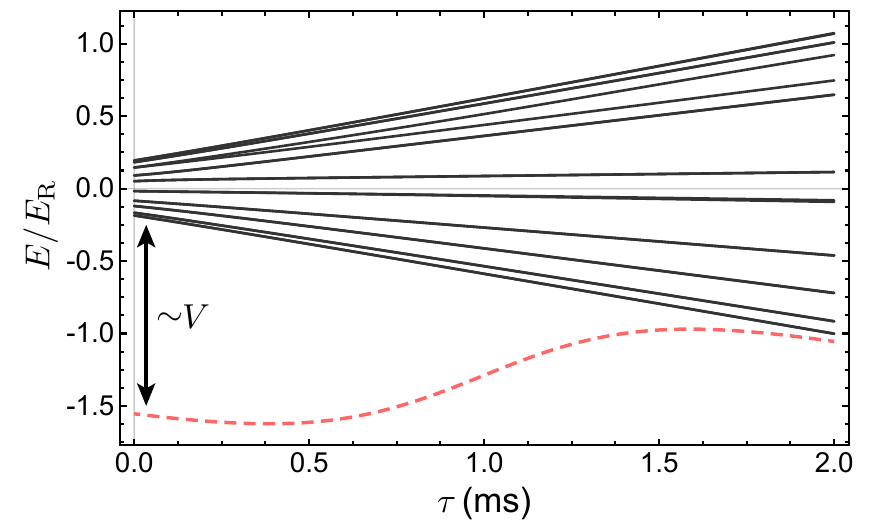}
	\caption{\label{FIG:figS2}
		\textbf{Ramp adiabaticity.}
		Noninteracting zigzag lattice eigenstate energies vs time $\tau$ during a ramp procedure to load the ground state, shown here for final disorder parameters $\Delta_\text{f}/t = 4$ and $\varphi = 0$.
		Red dashed curve shows the ground state, and solid black curves represent the $20$ excited states.
 	}
\end{figure}

One complication arises from the small spacing between applied spectral components (frequency teeth), as mentioned in the ``off-resonant excitations'' section above and shown in Fig.~\pref{FIG:figS1}{b}.
To implement the pseudorandom site energy shifts, we detune the frequencies away from Bragg resonances, thus moving some frequency teeth closer together and some further apart.
We restrict the maximum absolute detuning applied to any one frequency tooth to be less than half the spacing between adjacent teeth, or else it would then be more closely associated with a different Bragg resonance. While it would be preferable to not work at such extreme detunings, proportionally smaller tunneling strengths would be required to obtain the same final values of $\Delta/t$.
With the presence of our localized potential well, this restriction of the maximum detuning imposes the condition $(2\Delta+V)/\hbar < 2\pi\times 4056~\text{Hz}$, where the spacing between relevant first- and second-order Bragg transition frequencies is $4E_\text{R}/\hbar\approx 2\pi\times 8111~\text{Hz}$.
This means that there is some trade-off between the depth of our initial well and the initial disorder strength.
We choose to ramp up the disorder from $\Delta_0 = 0.2\Delta_f$ to a final disorder value $\Delta_f$ (for a range $1.0 \leq \Delta_f/t \leq 4.25$), and use the condition above to dictate the initial well depth $V$.

These ramps can be summarized in a plot of the eigenstate energies as a function of time during the ramps, shown in Fig.~\ref{FIG:figS2} for loading the ground state.
At time $\tau=0$, we begin with all tunnelings turned off, such that the states are separated based on their eigenstate energies.
The initial potential well of depth $V$ further separates the initially populated ground state (dashed red curve) at site $n=0$ from the rest of the spectrum.
As the ramp progresses, none of the states cross in energy, and by following the ground state (excited state) in a nearly adiabatic quasistatic fashion, we prepare a state that is close to the ground state (excited state) of the full Hamiltonian.
Then, by measuring the population distribution relating to this state in the synthetic lattice, we determine whether it remains localized or becomes delocalized for various $\Delta/t$ values.

We note that the procedure for the 1D Aubry-Andr\'{e} studies of Fig.~3 is similar to the one described here, but for $t'=0$, $\phi = 0$, and $V = 0$ (no additional well/hill) throughout. Also, we look not just at the extremal states, but instead probe various eigenstates.

\vsp
\pdfbookmark[1]{References}{References}
\hypersetup{bookmarksdepth=-2}


%